\documentclass[10pt,tightenlines,eqsecnum,floats,aps,amsmath,amssymb,nofootinbib,prd,showpacs]{revtex4}
\usepackage{color}
\usepackage{amsmath}
\usepackage{amsfonts}
\usepackage{times}
\usepackage{graphics}
\usepackage{bm}
\usepackage[colorlinks, linkcolor=blue, anchorcolor=blue, citecolor=red]{hyperref}
\usepackage{subeqnarray}
\usepackage{cases}
\usepackage{bbm}

\begin{document}

\title{Path Integral and Effective Hamiltonian in Loop Quantum Cosmology}
\author{Haiyun Huang}\email{haiyun@mail.bnu.edu.cn}
\author{Yongge Ma\footnote{Corresponding author}}\email{mayg@bnu.edu.cn}
\author{Li Qin}\email{qinli051@163.com}

\affiliation{Department of Physics, Beijing Normal University, Beijing 100875, China}

\begin{abstract}
We study the path integral formulation of Friedmann universe filled with a massless scalar field in loop quantum cosmology. All the isotropic models of $k=0,+1,-1$ are considered. To construct the path integrals in the timeless framework, a multiple group-averaging approach is proposed. Meanwhile, since the transition amplitude in the deparameterized framework can be expressed in terms of group-averaging, the path integrals can be formulated for both deparameterized and timeless frameworks. Their relation is clarified. It turns out that the effective Hamiltonian derived from the path integral in deparameterized framework is equivalent to the effective Hamiltonian constraint derived from the path integral in timeless framework, since they lead to same equations of motion. Moreover, the effective Hamiltonian constraints of above models derived in canonical theory are confirmed by the path integral formulation.


\end{abstract}

\maketitle

\section{Introduction}
 A core concept of general relativity is that physical theory should be background independent, which indicates that the gravity and matter evolve as a whole entity without anything pre-existed. This idea is being carried out in loop quantum gravity (LQG)\cite{Ro04,Th07,As04,Ma07}, which is a non-perturbative and background independent quantization of general relativity. The construction of LQG inspired the research on spin foam models (SFMs) \cite{Rovelli}, which are proposed as the path-integral formulation for LQG. In SFMs, the transition amplitude between physical quantum states is formulated as a sum over histories of all physically appropriate states. The heuristic picture is the following. One considers a 4-dimensional spacetime region M bounded by two spacelike 3-surfaces, which would correspond to the initial and final states for the path integral. The feature of SFMs is to employ the spin network states in LQG to represent the initial and final states. A certain path between the two boundaries is a quantum 4-geometry interpolated between the two spin networks. The interpolated spin foams can be constructed by considering the dual triangulation of M and coloring its surfaces with half integers and edges with suitable intertwiners. In order to obtain the physical inner product between the two boundary states, we have to sum over all possible triangulations and colorings \cite{Ro04}. This is certainly difficult to be achieved.

 On the other hand, the idea and technique of LQG is successfully carried out in the symmetry-reduced models, known as Loop Quantum Cosmology (LQC) \cite{Bojowald}. Since the infinite degrees of freedom of gravity have been reduced to finite ones by homogeneity (and isotropy), LQC provides a simple arena to test the ideas and constructions of the full LQG. Therefore, it is also desirable to test the idea and construction of SFMs by LQC models \cite{Henderson}. It was shown in Refs.\cite{spinfoam1,spinfoam2} that the transition amplitude in the deparameterized LQC of $k=0$ Friedmann universe equals the physical inner product in the timeless framework, and concrete evidence has been provided in support of the general paradigm underlying SFMs through the lens of LQC. How to achieve local spinfoam expansion in LQC was also studied in Refs. \cite{on the spinfoam, RV2}. Recently, the effective Hamiltonian constraint of $k=0$ LQC was derived from the path integral formulation in timeless framework in Ref.\cite{most recent}, and the effective measure introduces some conceptual subtleties in arriving at the WKB approximation. The purpose of this paper is to study the path integral formulation of LQC mimicking the construction of SFMs and following \cite{Henderson,spinfoam1,spinfoam2,RV2,most recent}. The new ingredient in this work is to formulate the path integrals by multiple group-averaging rather than to divide a single group-averaging. Besides the $k=0$ Friedmann-Rorberston-Walker (FRW) model, the $k=\pm1$ models will also be considered. It turns out that the physical inner product between boundary states, defined by group-averaging approach, can be interpreted as sum over all possible quantum states. The relation between the path integral formalism in timeless framework and that in deparameterized framework can be revealed by the group-averaging. Moreover, we will be able to derive the effective action and thus Hamiltonian of the above LQC models by the path integral formalism in timeless framework as well as deparameterized framework. The effective Hamiltonian coincide with those obtained by canonical approach. Hence the path integral and canonical formulations of LQC confirm each other. The effective Hamiltonian constraints of above models in canonical theory are also confirmed by the path integral formulation. Starting from the canonical LQC, we will derive the path integral formulation and the effective Hamiltonian of $k=0,+1,-1$ FRW models respectively in the following Sections 2, 3, and 4. To simplify the calculations, we will employ the simplified LQC treatments \cite{Robustness}. Note that, although the path integrals of $k=0$ FRW LQC have been studied in both deparameterized and timeless frameworks, we propose a new formulation by the multiple group-averaging approach.

\section{Path Integral Formulation of $k=0$ LQC}

The model that we are considering is the spatially flat $(k=0)$ FRW universe filled with a massless scalar field. In the kinematical setting, one has to introduce an elementary cell ${\cal V}$ in the spatial manifold and restricts all integrations to this cell. Fix a fiducial flat metric ${{}^o\!q}_{ab}$ and denote by $V_o$ the volume of ${\cal V}$ in
this geometry. The gravitational phase space variables ---the connections and the density weighted triads --- can be expressed as $ A_a^i = c\, V_o^{-(1/3)}\,\, {}^o\!\omega_a^i$ and $E^a_i = p\, V_o^{-(2/3)}\,\sqrt{{}^o\!q}\,\, {}^o\!e^a_i$, where $({}^o\!\omega_a^i, {}^o\!e^a_i)$ are a set of orthonormal co-triads and triads compatible with ${{}^o\!q}_{ab}$ and adapted to ${\cal V}$. $p$ is related to the scale factor $a$ via $|p|=V_o^{2/3}a^2$. The fundamental Poisson bracket is given by: $ \{c,\, p\} = {8\pi G\gamma}/{3} $, where $G$ is the Newton's constant and $\gamma$ the Barbero-Immirzi parameter. The gravitational part of the Hamiltonian constraint reads $C_{\mathrm{grav}} = -6 c^2\sqrt{|p|}/\gamma^2$.  To apply the area gap $\Delta$ of full LQG to LQC \cite{overview}, it is convenient to introduce variable $\bar{\mu}$ satisfying
\begin{align}
\label{f to mu}
\bar{\mu}^2~|p|=\Delta\equiv(4\sqrt{3}\pi\gamma)\ell_p^2,
\end{align}
where $\ell_p^2=G\hbar$, and new conjugate variables \cite{Robustness,CS,DMY,YDM2}:
\begin{align}
v:=2\sqrt{3}~\bar{\mu}^{-3},~~~~~b:=\bar{\mu}c.\label{v,b}
\end{align}
The new canonical pair satisfies $\{b,v\}=\frac{2}{\hbar}$. On the other hand, the matter phase space consists of canonical variables $\phi$ and $p_{\phi}$ which satisfy $\{\phi,p_{\phi}\}=1 $. To mimic LQG, the polymer-like representation is employed to quantize the gravity sector. The kinematical Hilbert space for gravity then reads $\mathcal{H}_{kin}^{grav}=L^2(\mathbb{R}_{\textrm{Bohr}},d\mu_H)$, where $\mathbb{R}_{\textrm{Bohr}}$ is the Bhor compactification of the real line and $d\mu_H$ is the Haar measure on it \cite{mathematical}. It turns out that the eigenstates of volume operator $\widehat{v}$, which are labeled by real number $v$, constitute an orthonomal basis in $\mathcal{H}_{kin}^{grav}$ as $\langle v_1| v_2\rangle=\delta_{v_1,v_2}$. For the scalar matter sector, one just uses the standard Schrodinger representation for its quantization, where the kinematical Hilbert space is $\mathcal{H}_{kin}^{matt}=L^2(\mathbb{R},d\phi)$. The total kinematical Hilbert Space of the system is a tensor product, $\mathcal{H}_{kin}=\mathcal{H}_{kin}^{grav}\otimes\mathcal{H}_{kin}^{matt}$, of the above two Hilbert spaces. As a totally constrained system, the dynamics of this model is reflected in the Hamiltonian constraint $C_\mathrm{grav}+C_\mathrm{matt}=0$. Quantum mechanically, physical states are those satisfying quantum constraint equation
\begin{align}
(\widehat{C}_\mathrm{grav}+\widehat{C}_\mathrm{matt})\Psi(v,\phi)=0.\label{orin constraint}
\end{align}
It is not difficult to write the above equation as \cite{improved dynamics}
\begin{align}
\widehat{C}\Psi(v,\phi)\equiv(\frac{\widehat{p}^2_{\phi}}{\hbar^2}-\widehat{\Theta})\Psi(v,\phi)=0. \label{constraint}
\end{align}
On one hand, Eq.\eqref{constraint} indicates that we can get physical states by group averaging kinematical states as \cite{on the spinfoam}
\begin{align}
\Psi_f(v,\phi)=\lim\limits_{\alpha_o\rightarrow\infty}\frac{1}{2\alpha_o}\int_{-\alpha_o}^{\alpha_o} d\alpha ~e^{i\alpha\widehat{C}}~f(v,\phi),
\end{align}
and thus the physical inner product of two states reads
\begin{align}
\langle~f|g~\rangle_{phy}=\langle \Psi_f|g~\rangle=\lim\limits_{\alpha_o\rightarrow\infty}\frac{1}{2\alpha_o}\int_{-\alpha_o}^{\alpha_o} d\alpha ~\langle f|e^{i\alpha\widehat{C}}|g\rangle.\label{innerproduct}
\end{align}
As is known, in timeless framework the transition amplitude equals to the physical inner product \cite{spinfoam1,spinfoam2}, i.e.,
\begin{align}
A_{tls}(v_f, \phi_f;~v_i,\phi_i)=\langle v_f, \phi_f|v_i,\phi_i\rangle_{phy}=\lim\limits_{\alpha_o\rightarrow\infty}\frac{1}{2\alpha_o}
\int_{-\alpha_o}^{\alpha_o}d\alpha\langle v_f,\phi_f|e^{i\alpha\widehat{C}}|v_i,\phi_i\rangle\label{amplitude}
\end{align}
On the other hand, Eq.\eqref{constraint} can also be written as
\begin{align}
\partial^2_\phi\Psi(v,\phi)+\widehat{\Theta}\Psi(v,\phi)=0.\label{Klein-Gorden}
\end{align}
The similarity between Eq.\eqref{Klein-Gorden} and Klein-Gorden equation suggests that one can regard $\phi$ as internal time, with respect to which gravitational field evolves.  In this deparameterized framework, we focus on positive frequency solutions, i.e., those satisfying
\begin{align}
-i\partial_\phi\Psi_+(v,\phi)=\widehat{\sqrt{\Theta}}\Psi_+(v,\phi)\equiv\widehat{H}\Psi_+(v,\phi).
\label{positive frequency}
\end{align}
The transition amplitude in deparameterized framework is then given by
\begin{align}
A_{dep}(v_f,\phi_f;~v_i,\phi_i)=\langle v_f|e^{i\widehat{H}(\phi_f-\phi_i)}|v_i\rangle,
\label{deparameterized amplitude1}
\end{align}
where $|v_i\rangle$ and $|v_f\rangle$ are eigenstates of volume operator in $\mathcal{H}_{kin}^{grav}$, and $\phi$ is the internal time.

Next we are to deal with transition amplitude under timeless work by starting from \eqref{amplitude}. In order to compute $\langle v_f, \phi_f|e^{i\alpha\widehat{C}}|v_i,\phi_i\rangle$, a straightforward way is to split the exponential into $N$ identical pieces and insert complete basis as in \cite{most recent}. However, since $\alpha$ is the group-averaging parameter which goes from $-\infty$ to $\infty$, it is unclear whether $\alpha$ could be treated as the time variable $t$ in non-relativistic quantum mechanics path integral. We thus consider alternative path integral formulation by group-averaging. To this end, we notice the following identity,
\begin{align}
&\lim\limits_{\alpha_o\rightarrow\infty}\frac{1}{2\alpha_o}\int_{-\alpha_o}^{\alpha_o} d\alpha ~e^{i\alpha\widehat{C}}|v,\phi\rangle\nonumber\\
=&\lim\limits_{\alpha'_o,\alpha_o\rightarrow\infty}\frac{1}{2\alpha'_o}\int_{-\alpha'_o}^{\alpha'_o} d\alpha' \frac{1}{2\alpha_o}\int_{\alpha_o}^{\alpha_o} d\alpha ~e^{i(\alpha' +\alpha )\widehat{C}}|v,\phi\rangle,
\end{align}
which could be generalized as
\begin{align}
&\lim\limits_{\alpha_o\rightarrow\infty}\frac{1}{2\alpha_o}\int_{-\alpha_o}^{\alpha_o}d\alpha ~e^{i\alpha\widehat{C}}|v,\phi\rangle\nonumber\\
=&\lim\limits_{\tilde\alpha_{{No}},...,\tilde\alpha_{{1o}}\rightarrow\infty}\frac{1}{2\tilde\alpha_{{No}}}
\int_{-\tilde\alpha_\emph{{No}}}^{\tilde\alpha_\emph{{No}}} d\tilde\alpha_N...\frac{1}{2\tilde\alpha_{{1o}}}\int_{-\tilde\alpha_{{1o}}}^{\tilde\alpha_{{1o}}} d\tilde\alpha_1 ~e^{i(\tilde\alpha_1 +...+\tilde\alpha_N )\widehat{C}}|v,\phi\rangle.
\end{align}
Let $\alpha=\sum\limits_{n=1}^N\tilde\alpha_n$. Then this equation indicates that we did not expect each piece of the exponential $e^{i\alpha\widehat{C}}$ to be identical. Rather, they could be independent of each other.
In order to trace the power for expansion, we let $\tilde\alpha_n=\epsilon\alpha_n$, where $\epsilon=\frac{1}{N}$. Then \eqref{amplitude} becomes
\begin{align}
&A_{tls}(v_f, \phi_f;~v_i,\phi_i)\nonumber\\
=&\lim\limits_{\tilde\alpha_\emph{{No}},...,\tilde\alpha_\emph{{1o}}\rightarrow\infty}\frac{1}{2\tilde\alpha_\emph{{No}}}
\int_{-\tilde\alpha_\emph{{No}}}^{\tilde\alpha_\emph{{No}}} d(\epsilon\alpha_N)...\frac{1}{2\tilde\alpha_\emph{{1o}}}\int_{-\tilde\alpha_\emph{{1o}}}^{\tilde\alpha_\emph{{1o}}} d(\epsilon\alpha_1) \langle v_f, \phi_f|e^{i\sum\limits_{n=1}^N{\epsilon\alpha_n}\widehat{C}}|v_i,\phi_i\rangle\nonumber\\
=&\lim\limits_{\alpha_\emph{{No}},...,\alpha_\emph{{1o}}\rightarrow\infty}\frac{1}{2\alpha_\emph{{No}}}
\int_{-\alpha_\emph{{No}}}^{\alpha_\emph{{No}}} d\alpha_N...\frac{1}{2\alpha_\emph{{1o}}}\int_{-\alpha_\emph{{1o}}}^{\alpha_\emph{{1o}}} d\alpha_1 \langle v_f, \phi_f|e^{i\sum\limits_{n=1}^N{\epsilon\alpha_n}\widehat{C}}|v_i,\phi_i\rangle,
\label{amplitude2}
\end{align}
where $\alpha_\emph{no}=\tilde\alpha_\emph{no}/\epsilon, n=1,2,..,N$. Next, we are going to insert a set of complete basis at each knot.
Notice that $|v,\phi\rangle$ is the eigenstate of both volume operator and scalar operator simultaneously in $\mathcal{H}_{kin}$, which can be written as $|v\rangle|\phi\rangle$ for short of $|v\rangle\otimes|\phi\rangle$,
and
\begin{align}
\mathbbm{1}_{kin}=\mathbbm{1}_{kin}^{grav}\otimes\mathbbm{1}_{kin}^{matt}=\sum\limits_{v}|v\rangle\langle v|\int d\phi~|\phi\rangle\langle\phi|.
\end{align}
Thus, we have
\begin{align}
\langle v_f, \phi_f|e^{i\sum\limits_{n=1}^N{\epsilon\alpha_n}\widehat{C}}|v_i,\phi_i\rangle=\sum\limits_{v_{N-1},...v_1}\int d\phi_{N-1}...d\phi_1\prod\limits_{n=1}^N\langle \phi_n|\langle v_n|e^{i\epsilon\alpha_n\widehat{C}}|v_{n-1}\rangle\phi_{n-1}\rangle,
\label{insert basis}
\end{align}
where $v_f=v_N,\phi_f=\phi_N,v_i=v_0,\phi_i=\phi_0$ have been set. Since the constraint operator $\widehat{C}$ has been separated into gravitational part and material part, which live in $\mathcal{H}_{kin}^{grav}$ and $\mathcal{H}_{kin}^{matt}$ separately, we could calculate the exponential on each kinematical space separately. For the material part, one gets
\begin{align}
&\langle{\phi_n}|e^{i\epsilon\alpha_n\frac{\widehat{p}^2_\phi}{\hbar^2}}|\phi_{n-1}\rangle\nonumber\\
=&\int dp_{\phi_n}\langle{\phi_n}|p_{\phi_n}\rangle\langle p_{\phi_n}|e^{i\epsilon\alpha_n\frac{\widehat{p}^2_\phi}{\hbar^2}}|\phi_{n-1}\rangle\nonumber\\
=&\frac{1}{2\pi\hbar}\int dp_{\phi_n}e^{i\epsilon(\frac{p_{\phi_n}}{\hbar}\frac{\phi_n-\phi_{n-1}}{\epsilon}
+\alpha_n\frac{{p}^2_{\phi_n}}{\hbar^2})}.
\label{material amplitude}
\end{align}
As for the gravitational part, in the limit $N\rightarrow\infty(\epsilon\rightarrow0)$, the operator $e^{-i\epsilon\alpha_n \widehat{\Theta}}$ can be expanded to the first order, and hence we get
\begin{align}
\langle v_{n}|e^{-i\epsilon\alpha_n \widehat{\Theta}}|v_{n-1}\rangle=\delta_{v_n,v_{n-1}}-i\epsilon\alpha_n\langle v_{n}|\widehat{\Theta}|v_{n-1}\rangle+\mathcal{O}(\epsilon^2).
\label{piece}
\end{align}
To simplify the calculation, we now employ the Hamiltonian operator $\widehat{\Theta}=\widehat{\Theta}_0$ in the simplified LQC of $k=0$ FRW model \cite{Robustness},  where $\widehat{\Theta}_0$ is a second order difference operator given by
\begin{align}
\widehat{\Theta}_0\Psi(v)=-\frac{3\pi G}{4}v
[(v+2)\Psi(v+4)-2v\Psi(v)+(v-2)\Psi(v-4)].\label{theta0}
\end{align}
This equation leads to
\begin{align}
\langle v_{n}|\widehat{\Theta}_0|v_{n-1}\rangle=-\frac{3\pi G}{4}v_{n-1}\frac{v_{n}+v_{n-1}}{2}
(\delta_{v_{n},v_{n-1}+4}-2\delta_{v_{n},v_{n-1}}+\delta_{v_{n},v_{n-1}-4}).
\label{matrix element}
\end{align}
Applying \eqref{matrix element} to \eqref{piece} and writing Kronecker delta as integral of $b_n$, which acts as the role of conjugate variable of $v_n$, we have
\begin{align}
&\langle v_{n}|e^{-i\epsilon\alpha_n \widehat{\Theta}_0}|v_{n-1}\rangle\nonumber\\
=&\frac{1}{\pi}\int^{\pi}_{0}db_n~e^{-ib_n(v_{n}-v_{n-1})/2}(1-i\alpha_n\epsilon\frac{3\pi G}{4}v_{n-1}\frac{v_{n}+v_{n-1}}{2}4\sin^2b_n)+\mathcal{O}(\epsilon^2).\nonumber\\
\label{gravitational amplitude}
\end{align}
Applying \eqref{material amplitude} and \eqref{gravitational amplitude} to \eqref{amplitude2}, and then taking `continuum limit', we obtain
\begin{align}
&A_{tls}(v_f, \phi_f;~v_i,\phi_i)\nonumber\\
=&\lim\limits_{N\rightarrow\infty}~~~~\lim\limits_{\alpha_\emph{{No}},...,\alpha_\emph{{1o}}\rightarrow\infty}
\left(\prod\limits_{n=1}^N\frac{1}{2\alpha_\emph{{no}}}\right)\int_{-\alpha_\emph{{No}}}^{\alpha_\emph{{No}}} d\alpha_N...\int_{-\alpha_\emph{{1o}}}^{\alpha_\emph{{1o}}} d\alpha_1\nonumber\\
&\times\int_{-\infty}^{\infty}d\phi_{N-1}...d\phi_1\left(\frac{1}{2\pi\hbar}\right)^N\int_{-\infty}^{\infty
}dp_{\phi_N}...dp_{\phi_1}~e^{i\epsilon(\frac{p_{\phi_n}}{\hbar}\frac{\phi_n-\phi_{n-1}}{\epsilon}
+\alpha_n\frac{{p}^2_{\phi_n}}{\hbar^2})}\nonumber\\
&\times\sum\limits_{v_{N-1},...,v_1}~~~\frac{1}{\pi^N}\int^{\pi}_{0}db_N...db_1~\prod
\limits_{n=1}^Ne^{ib_n(v_{n}-v_{n-1})/2}e^{-i\alpha_n\epsilon\frac{3\pi G}{4}v_{n-1}\frac{v_{n}+v_{n-1}}{2}4\sin^2b_n}\nonumber\\
=&\lim\limits_{N\rightarrow\infty}~~~~\lim\limits_{\alpha_\emph{{No}},...,\alpha_\emph{{1o}}\rightarrow\infty}
\left(\prod\limits_{n=1}^N\frac{1}{2\alpha_\emph{{no}}}\right)\int_{-\alpha_\emph{{No}}}^{\alpha_\emph{{No}}} d\alpha_N...\int_{-\alpha_\emph{{1o}}}^{\alpha_\emph{{1o}}} d\alpha_1\nonumber\\
&\times\int_{-\infty}^{\infty}d\phi_{N-1}...d\phi_1\left(\frac{1}{2\pi\hbar}\right)^N\int_{-\infty}^{\infty}
dp_{\phi_N}...dp_{\phi_1}\sum\limits_{v_{N-1},...,v_1}~~~\frac{1}{\pi^N}\int^{\pi}_{0}db_N...db_1\nonumber\\
&\times\prod\limits_{n=1}^{N}\exp{i\epsilon}\left[\frac{p_{\phi_n}}{\hbar}\frac{\phi_n-\phi_{n-1}}{\epsilon}
-\frac{b_n}{2}\frac{v_n-v_{n-1}}{\epsilon}+\alpha_n \left(\frac{p_{\phi_n}^2}{\hbar^2}-\frac{3\pi G}{4}v_{n-1}\frac{v_{n}+v_{n-1}}{2}4\sin^2b_n\right)\right].
\label{k=0 timeless}
\end{align}
Finally, we could write the above equation in path integral formulation as
\begin{align}
A_{tls}(v_f, \phi_f;~v_i,\phi_i)=c \int \mathcal{D}\alpha\int\mathcal{D}\phi\int\mathcal{D}p_{\phi}\int\mathcal{D}v\int\mathcal{D}b ~~\exp{\frac{i}{\hbar}\int_0^1d\tau~~ \left[p_\phi\dot\phi-\frac{\hbar}{2}b\dot{v}+{\hbar}{\alpha}(\frac{p_\phi^2}{\hbar^2}-3\pi Gv^2\sin^2b )\right]}\label{k=0H}
\end{align}
where $c$ is certain constant, and a dot over a letter denotes the derivative with respect to $\tau$. The effective Hamiltonian constraint can be read out from \eqref{k=0H} as
\begin{align}
C_{eff}=\frac{p_\phi^2}{\hbar^2}-3\pi Gv^2\sin^2b, \label{effC}
\end{align}
which coincides with that in Refs.\cite{DMY, most recent}.

On the other hand, for the deparameterized framework, we may still employ above group-averaging viewpoint. To this end, we define a new constraint operator $\widehat{C_+}=\frac{\widehat{p_{\phi}}}{\hbar}-\widehat{H}$. Then Eq.\eqref{positive frequency} can be rewritten as
\begin{align}
\widehat{C_+}\Psi_+(v,\phi)=0.
\end{align}
The transition amplitude for this new constraint reads
\begin{align}
A_{dep}(v_f,\phi_f;~v_i,\phi_i)=\lim\limits_{\alpha_o\rightarrow\infty}\frac{1}{2\alpha_o}\int_{-\alpha_o}^{\alpha_o} d\alpha \langle v_f,\phi_f|e^{i\alpha \widehat{C_+}}|v_i,\phi_i\rangle=\lim\limits_{\alpha_o\rightarrow\infty}\frac{1}{2\alpha_o}\int_{-\alpha_o}^{\alpha_o} d\alpha \langle v_f,\phi_f|2\widehat{|{p}_\phi|}\widehat{\theta(p_{\phi})}e^{i\alpha \widehat{C}}|v_i,\phi_i\rangle,
\label{deparameterized amplitude2}
\end{align}
where
\begin{align}
\widehat{|{p}_\phi|}|p_{\phi}\rangle=|p_{\phi}||p_{\phi}\rangle,~~
\widehat{\theta(p_{\phi})}|p_{\phi}\rangle=
\begin{cases}
0& \text{$p_{\phi}\leq0$}\\
|p_{\phi}\rangle& \text{$p_{\phi}>0$}.
\end{cases}
\end{align}
Similar to the timeless case, the integration over single $\alpha$ can be written as multiple integrations,
\begin{align}
&A_{dep}(v_f, \phi_f;~v_i,\phi_i)\nonumber\\
=&\lim\limits_{\alpha_\emph{{No}},...,\alpha_\emph{{1o}}\rightarrow\infty}\frac{1}{2\alpha_\emph{{No}}}
\int_{-\alpha_\emph{{No}}}^{\alpha_\emph{{No}}} d\alpha_N...\frac{1}{2\alpha_\emph{{1o}}}\int_{-\alpha_\emph{{1o}}}^{\alpha_\emph{{1o}}} d\alpha_1 ~~\langle v_f, \phi_f|e^{i\sum\limits_{n=1}^N{\epsilon\alpha_n}\widehat{C_+}}|v_i,\phi_i\rangle.
\label{deparameterized amplitude3}
\end{align}
Since we work in the deparameterized case, it is reasonable to insert in the completeness relation \cite{two points,sumover}
\begin{align}
\mathbbm{1}=\sum\limits_v|v,\phi,+\rangle\langle v,\phi,+|,
\end{align}
where
\begin{align}
|v,\phi,+\rangle=\lim\limits_{\beta_o\rightarrow\infty}\frac{1}{2\beta_o}\int_{-\beta_o}^{\beta_o} d\beta~e^{i\beta\widehat{C_+}}|v,\phi\rangle.\label{basis}
\end{align}
Note that the Hilbert space in deparameterized framework is unitarily equivalent to the physical Hilbert space with the basis \eqref{basis}.
Then the first piece of the exponential in \eqref{deparameterized amplitude3} becomes
\begin{align}
&\lim\limits_{\alpha_1,\beta_{1o}\rightarrow\infty}\frac{1}{2\alpha_{1o}}\int_{-\alpha_{1o}}^{\alpha_{1o}} d\alpha_1\frac{1}{2\beta_{1o}}\int_{-\beta_{1o}}^{\beta_{1o}} d\beta_1\langle v_1,\phi_1|e^{-i\beta_1\widehat{C_+}}e^{i\epsilon\alpha_1\widehat{C_+}}|v_{0},\phi_{0}\rangle\nonumber\\
=&\lim\limits_{\alpha_1\rightarrow\infty}\frac{1}{2\alpha_{1o}}\int_{-\alpha_{1o}}^{\alpha_{1o}} d\alpha_1\langle v_1,\phi_1|e^{i\epsilon\alpha_1\widehat{C_+}}|v_{0},\phi_{0}\rangle,
\end{align}
and the last piece of the exponential reads
\begin{align}
&\lim\limits_{\beta_{No}\rightarrow\infty}\frac{1}{2\alpha_{No}}\int_{-\alpha_{No}}^{\alpha_{No}} d\alpha_N\frac{1}{2\beta'_{1o}}\int_{-\beta'_{No}}^{\beta'_{No}} d\beta'_N\langle v_N,\phi_N|e^{i\epsilon\alpha_N\widehat{C_+}}e^{i\beta'_N\widehat{C_+}}|v_{N-1},\phi_{N-1}\rangle\nonumber\\
=&\lim\limits_{\alpha_N\rightarrow\infty}\frac{1}{2\alpha_{No}}\int_{-\alpha_{No}}^{\alpha_{No}} d\alpha_N\langle v_N,\phi_N|e^{i\epsilon\alpha_N\widehat{C_+}}|v_{N-1},\phi_{N-1}\rangle.
\end{align}
The remaining pieces of the exponential can also be expressed as
\begin{align}
&\lim\limits_{\alpha_{no},\beta_{no},\beta'_{no}\rightarrow\infty}\frac{1}{2\alpha_{no}}\int_{-\alpha_{no}}^{\alpha_{no}} d\alpha_n\frac{1}{2\beta_{no}}\int_{-\beta_{no}}^{\beta_{no}} d\beta_n\frac{1}{2\beta'_{no}}\int_{-\beta'_{no}}^{\beta'_{no}} d\beta'_n~\langle v_n, \phi_n|e^{-i\beta_n\widehat{C_+}}e^{i\epsilon\alpha_n\widehat{C_+}}e^{i\beta'_n\widehat{C_+}}|v_{n-1},\phi_{n-1}
\rangle\nonumber\\
=&\lim\limits_{\alpha_n\rightarrow\infty}\frac{1}{2\alpha_{no}}\int_{-\alpha_{no}}^{\alpha_{no}} d\alpha_n\langle v_n,\phi_n|e^{i\epsilon\alpha_n\widehat{C_+}}|v_{n-1},\phi_{n-1}\rangle.
\end{align}
So \eqref{deparameterized amplitude3} becomes
\begin{align}
&A_{dep}(v_f, \phi_f;~v_i,\phi_i)\nonumber\\
=&\lim\limits_{\alpha_{{No}},...,\alpha_{{1o}}\rightarrow\infty}\frac{1}{2\alpha_{{No}}}
\int_{-\alpha_{{No}}}^{\alpha_{{No}}} d\alpha_N...\frac{1}{2\alpha_{{1o}}}\int_{-\alpha_{{1o}}}^{\alpha_{{1o}}} d\alpha_1\sum\limits_{v_{N-1},...,v_1}\prod\limits_{n=1}^N~\langle v_n, \phi_n|e^{i\epsilon\alpha_n\widehat{C_+}}|v_{n-1},\phi_{n-1}\rangle.\nonumber\\
=&\lim\limits_{\alpha_\emph{{No}},...,\alpha_\emph{{1o}}\rightarrow\infty}\frac{1}{2\alpha_\emph{{No}}}
\int_{-\alpha_\emph{{No}}}^{\alpha_\emph{{No}}} d\alpha_N...\frac{1}{2\alpha_\emph{{1o}}}\int_{-\alpha_\emph{{1o}}}^{\alpha_\emph{{1o}}} d\alpha_1\sum\limits_{v_{N-1},...,v_1}\prod\limits_{n=1}^N~\langle v_n, \phi_n|2\widehat{|{p}_{\phi_n}|}\widehat{\theta({p}_{\phi_n})}e^{i\epsilon\alpha_n\widehat{C}}|v_{n-1},\phi_{n-1}
\rangle,\label{damplitude}
\end{align}
where \eqref{deparameterized amplitude2} is applied in second step.
Now we can split each piece in \eqref{damplitude} into gravitational and material parts. Calculations similar to those in timeless framework lead to
\begin{align}
&A_{dep}(v_f, \phi_f;~v_i,\phi_i)\nonumber\\
=&\lim\limits_{N\rightarrow\infty}~~~~\lim\limits_{\alpha_\emph{{No}},...,\alpha_\emph{{1o}}\rightarrow\infty}
\frac{1}{2\alpha_\emph{{No}}}\int_{-\alpha_\emph{{No}}}^{\alpha_\emph{{No}}} d\alpha_N...\frac{1}{2\alpha_\emph{{1o}}}\int_{-\alpha_\emph{{1o}}}^{\alpha_\emph{{1o}}} d\alpha_1\left(\frac{1}{2\pi\hbar}\right)^N\int_{-\infty}^{\infty}dp_{\phi_N}...dp_{\phi_1}\sum
\limits_{v_{N-1},...,v_1}~~~\frac{1}{\pi^N}\int^{\pi}_{0}db_N...db_1\nonumber\\
&\times\prod\limits_{n=1}^{N}2|p_{\phi_n}|\theta(p_{\phi_n})\exp{i\epsilon}\left[\frac{p_{\phi_n}}{\hbar}
\frac{\phi_n-\phi_{n-1}}{\epsilon}-\frac{b_n}{2}\frac{v_n-v_{n-1}}{\epsilon}+\alpha_n \left(\frac{p_{\phi_n}^2}{\hbar^2}-\frac{3\pi G}{4}v_{n-1}\frac{v_{n}+v_{n-1}}{2}4\sin^2b_n\right)\right].
\end{align}
We can integrate out $\alpha_n$ and $p_{\phi_n}$ and arrive at
\begin{align}
&A_{dep}(v_f, \phi_f;~v_i,\phi_i)\nonumber\\
=&\lim\limits_{N\rightarrow\infty}~~~\lim\limits_{\alpha_\emph{{No}},...,\alpha_\emph{{1o}}\rightarrow\infty}
\left(\prod\limits_{n=1}^N\frac{1}{2\alpha_\emph{{no}}}\right)
\epsilon^{-N}\left(\frac{1}{2\pi\hbar}\right)^N\hbar^N\sum\limits_{v_{N-1},...,v_1}~~~\frac{1}{\pi^N}
\int^{\pi}_{0}db_N...db_1\nonumber\\
&\times\prod\limits_{n=1}^{N}\exp{i\epsilon(\phi_f-\phi_i)}\left[\sqrt{\frac{3\pi G}{4}v_{n-1}\frac{v_{n}+v_{n-1}}{2}4\sin^2b_n}-\frac{b_n}{2}\frac{v_n-v_{n-1}}{\epsilon(\phi_f-\phi_i)}\right]\nonumber\\
=&c'\int\mathcal{D}v\int\mathcal{D}b~\exp\frac{i}{\hbar}\int d\phi(\sqrt{3\pi G}\hbar v\sin b
-\frac{\hbar\dot{v}b}{2}),\label{damplitude2}
\end{align}
where $\dot{v}=\frac{dv}{d\phi}$. The effective Hamiltonian in deparameterized framework can be read out from \eqref{damplitude2} as $H_{eff}=\sqrt{3\pi G}\hbar v\sin b$, which is just the deparameterized Hamiltonian of \eqref{effC}.

Let us summarize what we have done to construct the path integral formulation of k=0 LQC. First, the transition
amplitude is expressed in terms of multiple group averaging. Secondly, we insert in complete basis and calculate each exponential to the second order. Thirdly, by taking `continuum limit', the path integral formulation under timeless framework is derived. This technique could also be applied to the path integral calculation of other constrained systems. Finally, by reexpressing the deparameterized equation in the form of constraint, the path integral of deparameterized framework is also formulated in this approach.
Hence we have got path integral formulation of timeless framework as well as deparameterized framework. The relation of these two kinds of path integral formulation is revealed by \eqref{deparameterized amplitude2}. They are in accordance with each other in the sense of effective theory.

\section{Path Integral Formulation of $k=1$ LQC}

The spatial manifold $M$ of $k=1$ FRW model is topologically a 3-sphere $\mathbb{S}^3$. Unlike $k=0$ model, we do not have to introduce a fiducial cell in it. Rather, the volume $V_o$ of $M$ with reference to a fiducial metric $^oq_{ab}$ is finite. We shall set $\ell_o := V_o ^{1/3}$. The gravitational connections and the density weighted triads can still be expressed as $ A_a^i = c\, V_o^{-(1/3)}\,\,
{}^o\!\omega_a^i$ and $E^a_i = p\,
V_o^{-(2/3)}\,\sqrt{{}^o\!q}\,\, {}^o\!e^a_i$. The fundamental Poisson bracket remains $
\{c,\, p\} = {8\pi G\gamma}/{3} $, and we introduce the same new conjugate variables  as in \eqref{v,b}. However, the gravitational part of the Hamiltonian constraint of $k=1$ model is different from that in $k=0$ model due to non-vanishing spatial curvature. In quantum theory, we have the same Hilbert space $\mathcal{H}_{kin}=\mathcal{H}_{kin}^{grav}\otimes\mathcal{H}_{kin}^{matt}$. The gravitational Hamiltonian constraint operator $\widehat{\Theta}_1$ can be written as a combination of diagonal and off-diagonal part, whose action reads \cite{k=1}
\begin{align}
{\widehat{\Theta}_1}\Psi(v)={e^{-i\ell_0f}\widehat{\Theta}_0e^{i\ell_0f}}\Psi(v)+{\Tilde{\Theta}_1}\Psi(v),
\end{align}
where
\begin{align}
&f=\frac{sgn~v}{4}|\frac{v}{K}|^{\frac{2}{3}}=f(v), ~~K=\frac43\sqrt{\frac{\pi\gamma\ell_p^2}{3\Delta}}
,\nonumber\\
&{\Tilde{\Theta}_1}
=3\pi Gv^2[-\sin^2(\frac{\bar{\mu}\ell_0}{2})+(1+\gamma^2)(\frac{\bar{\mu}\ell_0}{2})^2]\equiv\Tilde{\Theta}_1(v),
\end{align}
and $\widehat{\Theta}_0\equiv\widehat{\Theta}$ is the Hamiltonian operator in $k=0$ model. The definition of $K$ is in accordance with \cite{YDM2}. For simplicity, we still employ the simplified  treatment and use Eq.\eqref{theta0}.
To calculate the transition amplitude between two basis vectors in the timeless framework as \eqref{amplitude}, we also need to calculate Eq.\eqref{piece} with $\widehat{\Theta}$ replaced by $\widehat{\Theta}_1$. Note that the diagonal part $\langle v_{n}|\tilde{\Theta}_1|v_{n-1}\rangle$ is easy to deal with by writing delta function as an integration over $b$. We are going to evaluate the nontrivial off-diagonal part,
\begin{align}
\langle &v_{n}|{e^{-i\ell_0f}\widehat{\Theta}_0e^{i\ell_0f}}|v_{n-1}\rangle\nonumber\\
=-&\frac{3\pi G}{4}v_{n-1}\frac{v_{n}+v_{n-1}}{2}
\left(e^{-i\ell_0[f(v_{n-1}+4)-f(v_{n-1})]}\delta_{v_{n},v_{n-1}+4}-2\delta_{v_{n},v_{n-1}}
+e^{-i\ell_0[f(v_{n-1}-4)-f(v_{n-1})]}\delta_{v_{n},v_{n-1}-4}\right).
\label{off diagnal}
\end{align}
We assume $v>4$ and do Taylor expansion as
\begin{align}
&f(v+4)-f(v)=\frac{v^{2/3}}{4K}\left(\frac23\frac4v+\frac1{2!}\frac23(-\frac13)(\frac4v)^2
+\frac1{3!}\frac23(-\frac13)(-\frac43)(\frac4v)^3+...\right)\equiv g(v)+h(v),\nonumber\\
&f(v-4)-f(v)=\frac{v^{2/3}}{4K}\left(-\frac23\frac4v+\frac1{2!}\frac23(-\frac13)(\frac4v)^2
-\frac1{3!}\frac23(-\frac13)(-\frac43)(\frac4v)^3+...\right)\equiv -g(v)+h(v),
\label{diagonal part}
\end{align}
where $h(v)$ denotes sum of the odd terms and $g(v)$ that of the evens.
Thus, Eq.(\ref{off diagnal}) turns out to be
\begin{align}
\langle &v_{n}|{e^{-i\ell_0f}\widehat{\Theta}_0e^{i\ell_0f}}|v_{n-1}\rangle\nonumber\\
=&\frac{3\pi G}{4}~v_{n-1}\frac{v_{n}+v_{n-1}}{2}\frac{1}{\pi}\int^{\pi}_{0} db_n~e^{-ib_n(v_{n}-v_{n-1})/2}~\left[e^{-i\ell_0h(v_{n-1})}2\cos(2b_n-g(v_{n-1})\ell_0)-2\right].
\end{align}
If we denote $\Lambda_{n}=e^{-i\ell_0h(v_{n-1})}2\cos(2b_n-g(v_{n-1})\ell_0)-2$, the matrix element of $\widehat{\Theta}_1$ can be calculated as
\begin{align}
&~~~~\langle v_{n}|\widehat{\Theta}_1|v_{n-1}\rangle\nonumber\\
&=\frac{1}{\pi}\int^{\pi}_{0} db_n~e^{-ib_n(v_{n}-v_{n-1})/2}\left(\Tilde{\Theta}_1(v_{n-1})+{\frac{3\pi G}{4}}{}~v_{n-1}\frac{v_{n}+v_{n-1}}{2}\Lambda_n\right).
\end{align}
Therefore we have
\begin{align}
&\langle v_{n}|e^{-i\epsilon\alpha \widehat{\Theta}_1}|v_{n-1}\rangle\nonumber\\
=&\frac{1}{\pi}\int^{\pi}_{0}db_n~e^{-ib_n(v_{n}-v_{n-1})/2}\left[1-i\alpha\epsilon\left(\Tilde{\Theta}_1(v_{n-1})
+{\frac{3\pi G}{4}}{}~v_{n-1}\frac{v_{n}+v_{n-1}}{2}\Lambda_n\right)\right]+\mathcal{O}(\epsilon^2).
\label{k=1 exponential}
\end{align}
Replacing \eqref{gravitational amplitude} with \eqref{k=1 exponential} and following the procedure as in $k=0$ model, finally we arrive at the path integral formulation of $k=1$ FRW model under timeless framework as
\begin{align}
&A_{tls}(v_f,\phi_f;v_i,\phi_i)^{k=1}\nonumber\\
=&c_1\int \mathcal{D}\alpha\int\mathcal{D}\phi\int\mathcal{D}p_{\phi}\int\mathcal{D}v\int\mathcal{D}b~\exp\frac{i}{\hbar}\int_0^1 d\tau\left[p_{\phi}\dot{\phi}-\frac{\hbar\dot{v}b}{2}+\hbar\alpha\left(\frac{p_{\phi}^2}{\hbar}-\Tilde{\Theta}_1(v)
-{\frac{3\pi G}{4}}{}~v^2\Lambda(v,b)\right)\right].\label{tamplitude+}
\end{align}
 However, since $\Lambda(v,b)$ is a complex function, \eqref{tamplitude+} would imply a complex effective action or Hamiltonian. To avoid this problem, we notice that in semiclassical regime, the higher power terms in \eqref{diagonal part} decrease dramatically as $v$ increases, and hence only the first term, denoted by $g_1$, dominates. For example, when $v=100$, $\frac{f(v+4)-f(v)-g_1(v)}{f(v+4)-f(v)}<1\%$. As a matter of fact, in the simplified LQC treatment a similar approximation has already been used. In this sense, we could reserve only $g_1$ term and get
 \begin{align}
 \Lambda(v,b)=2\left[\cos(2b-g_1(v)\ell_0)-1\right]=4\sin^2(b-\bar{\mu}(v)\ell_0/2).
 \end{align}
Consequently, the transition amplitude turns out to be
\begin{align}
&A_{tls}(v_f,\phi_f;v_i,\phi_i)^{k=1}\nonumber\\
=&c_1\int \mathcal{D}\alpha\int\mathcal{D}\phi\int\mathcal{D}p_{\phi}\int\mathcal{D}v\int\mathcal{D}b~\exp\frac{i}{\hbar}\int_0^1 d\tau\left[p_{\phi}\dot{\phi}-\frac{\hbar\dot{v}b}{2}+\hbar\alpha\left(\frac{p_{\phi}^2}{\hbar}-\Tilde{\Theta}_1(v)
-{3\pi G}~v^2\sin^2(b-\bar{\mu}(v)\frac{\ell_0}2)\right)\right]
\end{align}
where a dot over a letter denotes the derivative with respect to $\tau$.
From this formulation we could read out the effective Hamiltonian constraint of timeless framework as
\begin{align}
\label{eff Hamil timeless}
C_\emph{eff}^{\textit{tls}}=\frac{p_{\phi}^2}{\hbar}-\Tilde{\Theta}_1(v)-{3\pi G}{}~v^2\sin^2(b-\bar{\mu}(v)\frac{\ell_0}{2}).
\end{align}
Note that in \cite{k=1} an effective Hamiltonian constraint under timeless framework was also given by canonical theory as
\begin{align}
\mathcal{H}_{\emph{eff}}&:=\frac{\mathcal{C}_{\emph{eff}}}{16\pi G}=\frac{A(v)}{16\pi G}\left[\sin^2\bar{\mu}(c-\frac{\ell_0}{2})-\sin^2(\frac{\bar{\mu}\ell_0}{2})+(1+\gamma^2)
\frac{\bar{\mu}^2\ell_0^2}{4}\right]+(\frac{8\pi\gamma\ell_{pl}^2}{6})^{-\frac{3}{2}}B(v)
\frac{p_{\phi}^2}{2},
\label{effective constraint}
\end{align}
where
\begin{align}
{A}(v)&=-\frac{27K}{4}\sqrt{\frac{8\pi}{6}}\frac{\ell_{p}}{\gamma^{3/2}}|v|\left||v-1|-|v+1|\right|,\nonumber\\
B(v)&=\left(\frac{3}{2}\right)^3K|v|\left||v+1|^{1/3}-|v-1|^{1/3}\right|^3.
\label{A,B}
\end{align}
In large $v$ regime, we have
\begin{align}
A(v)&=-\frac{27K}{4}\sqrt{\frac{8\pi}{6}}\frac{\ell_{p}}{\gamma^{3/2}}2v,\nonumber\\
B(v)&=\frac{K}{v}.
\end{align}
Then the effective Hamiltonian constraint (\ref{effective constraint}) can be simplified as
\begin{align}
\label{eff Hamiltonian sim}
\mathcal{H}_\emph{eff}=\frac{\frac{p_{\phi}^2}{\hbar^2}-\{\Tilde{\Theta}_1(v)+{3\pi G}{}~v^2\sin^2[b-\bar{\mu}(v)\frac{\ell_0}{2}]\}}{2\frac{v}{K\hbar^2}(\frac{8\pi\gamma\ell_{pl}^2}{6})^{\frac{3}{2}}}.
\end{align}
Note that \eqref{eff Hamiltonian sim} is slightly different from \eqref{eff Hamil timeless} by a factor $\frac12\frac{K\hbar^2}{v}(\frac{8\pi\gamma\ell_{pl}^2}{6})^{-\frac{3}{2}}$, which will not affect effective equations of motion. The appearance of this factor is due to the difference of Eq.\eqref{orin constraint} from Eq.\eqref{constraint}.

Taking into account of $\mathcal{H}_{\emph{eff}}=0$, the effective canonical equations can be derived from \eqref{eff Hamiltonian sim} as
\begin{subequations}
\label{equations from Hamiltonian}
\begin{align}
\frac{dv}{d\tau}&=\{v,\mathcal{H}_{\emph{eff}}\}=\frac{3\pi Gv^2\sin2[b-\bar{\mu}(v)\frac{\ell_0}{2}]}{\frac{v}{K\hbar^2}(\frac{8\pi\gamma\ell_{pl}^2}{6})^{\frac{3}{2}}},\\
\frac{db}{d\tau}&=\{b,\mathcal{H}_{\emph{eff}}\}=-\frac{\Tilde{\Theta}'_1(v)+{6\pi G}{}~v\sin^2[b-\bar{\mu}(v)\frac{\ell_0}{2}]-3\pi Gv^2\sin2[b-\bar{\mu}(v)\frac{\ell_0}{2}]\bar{\mu}'(v)}{\frac{v}{K\hbar^2}(\frac{8\pi\gamma
\ell_{pl}^2}{6})^{\frac{3}{2}}},\\
\frac{d\phi}{d\tau}&=\{\phi,\mathcal{H}_{\emph{eff}}\}=\frac{2p_{\phi}}{2\frac{v}{K\hbar^2}(\frac{8\pi\gamma
\ell_{pl}^2}{6})^{\frac{3}{2}}}=\frac{\sqrt{\Tilde{\Theta}_1(v)+{3\pi G}{}~v^2\sin^2[b-\bar{\mu}(v)\frac{\ell_0}{2}]}}{\frac{v}{K\hbar^2}(\frac{8\pi\gamma
\ell_{pl}^2}{6})^{\frac{3}{2}}},\label{dphidtau}
\end{align}
\end{subequations}
where a prime on a letter denotes the derivative with respect to the argument.
Note that in order to compare the effective equations with those in deparameterized framework, we focused on `positive frequency' as Eq.\eqref{positive frequency} and thus chose the plus sign in front of the square-root in Eq.\eqref{dphidtau}. This means that the internal time $\phi$ is monotonically increasing with $\tau$.

Following the same procedure in $k=0$ model, we also obtain the path integral formulation of $k=1$ model under deparameterized framework as
\begin{align}
&A_{dep}(v_f,\phi_f;v_i,\phi_i)^{k=1}\nonumber\\
=&c'_1\int\mathcal{D}v\int\mathcal{D}b~\exp\frac{i}{\hbar}\int d\phi(\sqrt{\Tilde{\Theta}_1(v)+{3\pi G}{}~v^2\sin^2[b-\bar{\mu}(v)\frac{\ell_0}{2}]}\hbar-\frac{\hbar\dot{v}b}{2}).
\end{align}
This enable us to read out an effective action as
\begin{align}
S[b,v]&=\int d\phi(\sqrt{\Tilde{\Theta}_1(v)+{3\pi G}{}~v^2\sin^2[b-\bar{\mu}(v)\frac{\ell_0}{2}]}\hbar-\frac{\hbar\dot{v}b}{2}),
\label{eff action 1}
\end{align}
and thus the effective Hamiltonian under deparameterized framework,
\begin{align}
\mathcal{H}_{\textit{eff}}^{dep}=-\sqrt{\Tilde{\Theta}_1(v)+{3\pi G}{}~v^2\sin^2[b-\bar{\mu}(v)\frac{\ell_0}{2}]}\hbar.
\end{align}
Hence the effective equations of this model can be derived by following Poisson brackets
\begin{subequations}
\begin{align}{}
\frac{dv}{d\phi}&=\{v,\mathcal{H}_{\textit{eff}}^{dep}\}=\frac{3\pi Gv^2\sin2[b-\bar{\mu}(v)\frac{\ell_0}{2}]}{\sqrt{\Tilde{\Theta}_1(v)+{3\pi G}{}~v^2\sin^2[b-\bar{\mu}(v)\frac{\ell_0}{2}]}
},\\
\frac{db}{d\phi}&=\{b,\mathcal{H}_{\textit{eff}}^{dep}\}=-\frac{\Tilde{\Theta}'_1(v)+{6\pi G}{}~v\sin^2[b-\bar{\mu}(v)\frac{\ell_0}{2}]-3\pi Gv^2\sin2[b-\bar{\mu}(v)\frac{\ell_0}{2}]\bar{\mu}'(v)}{\sqrt{\Tilde{\Theta}_1(v)+{3\pi G}{}~v^2\sin^2[b-\bar{\mu}(v)\frac{\ell_0}{2}]}},
\end{align}
\label{eff eqs from action}
\end{subequations}
It is easy to see that the effective equations \eqref{eff eqs from action} can be derived from equations \eqref{equations from Hamiltonian} by $\frac{dv}{d\phi}=\frac{dv/d\tau}{d\phi/d\tau}$ and $\frac{db}{d\phi}=\frac{db/d\tau}{d\phi/d\tau}$.  Therefore, the effective action in \eqref{eff action 1} predicts the same equations of motion for the $k=1$ FRW universe as those in \cite{k=1}. The quantum bounce driven by the quantum gravity effect would lead to a cyclic universe for this model. Moreover, the comparison between the path integrals in the deparameterized framework and timeless framework implies their equivalence in the sense of effective theory.

\section{Path Integral Formulation of $k=-1$ LQC}

In $k=-1$ FRW model, the spatial manifold is noncompact just as that of $k=0$ model. So we should still choose a fiducial cell $\cal{V}$ in the spatial manifold and restrict all integrations on it. The gravitational phase space variables $(A^i_a, E^a_i)$ and reduced new conjugate variables $(v,b)$ are defined in the same ways as $k=0$ model. In $k=-1$ LQC \cite{k=-1}, by the simplified treatment the off diagonal part of the gravitational constraint operator $\widehat{\Theta}_{-1}$ is exactly the Hamiltonian operator in $k=0$ model, while the $k=-1$ model contributes only diagonal part, i.e.
\begin{align}
\widehat{\Theta}_{-1}=\widehat{\Theta}_{0}+\tilde{\Theta}_{-1},\label{theta -1}
\end{align}
where
\begin{align}
\tilde{\Theta}_{-1}&=-B(v)^{-1}\frac{8\pi G\gamma^2V_0^{2/3}}{12K^{1/3}}|v|^{1/3}||v+1|-|v-1||.
\end{align}
In large $v$ regime, we have
\begin{align}
\tilde{\Theta}_{-1}(v)=-3\pi Gv^2\gamma^2V_0^{\frac{2}{3}}\bar{\mu}^2.
\end{align}
In the simplified treatment, we can still apply the $\widehat{\Theta}_0$ in \eqref{theta0} to \eqref{theta -1}. Thus, by Eq.\eqref{matrix element}, we can easily write down the matrix element of $\widehat{\Theta}_{-1}$ as
\begin{align}
&~~~~\langle v_{n}|\widehat{\Theta}_{-1}|v_{n-1}\rangle\nonumber\\
&=\frac{1}{\pi}\int^{\pi}_{0} db_n~e^{-ib_n(v_{n}-v_{n-1})/2}[\Tilde{\Theta}_{-1}(v_{n-1})+{3\pi G}{}~v_{n-1}\frac{v_{n}+v_{n-1}}{2}\sin^2\bar{\mu}c].
\label{k=-1}
\end{align}
Hence, through the approach outlined in Section 2, we can also obtain the path integral formulation of timeless framework as
\begin{align}
&A_{tls}(v_f,\phi_f;v_i,\phi_i)^{k=-1}\nonumber\\
=&c_{-1}\int \mathcal{D}\alpha\int\mathcal{D}\phi\int\mathcal{D}p_{\phi}\int\mathcal{D}v\int\mathcal{D}b~\exp\frac{i}{\hbar}\int_0^1 d\tau\left[p_{\phi}\dot{\phi}-\frac{\hbar\dot{v}b}{2}+\hbar\alpha\left(\frac{p_{\phi}^2}{\hbar}-{3\pi G}v^2\left(\sin^2b-\gamma^2V_0^{\frac{2}{3}}\bar{\mu}^2\right)\right)\right]
\end{align}
where a dot over a letter denotes the derivative with respect to $\tau$. The effective Hamiltonian constraint for $k=-1$ FRW model is then derived as
\begin{align}
\mathcal{H}_\emph{eff}^{\textit{tls}}=\frac{p_{\phi}^2}{\hbar}-{3\pi G}v^2\left(\sin^2b-\gamma^2V_0^{\frac{2}{3}}\bar{\mu}^2\right).
\label{tls H}
\end{align}
Also, Eq.\eqref{tls H} is in accordance with the Hamiltonian constraint given in  \cite{k=-1}, which reads
\begin{align}
\mathcal{H}_{\emph{eff}}&=-\frac{3\sqrt{|p|}}{8\pi G\gamma^2\bar{\mu}^2}\sin^2(\bar{\mu}c)+\frac{3\sqrt{|p|}V_0^{2/3}}{8\pi G}+|p|^{3/2}\rho_M\nonumber\\
&=\frac{\frac{p_{\phi}^2}{\hbar}-{3\pi G}v^2\left(\sin^2b-\gamma^2V_0^{\frac{2}{3}}\bar{\mu}^2\right)}{2\frac{v}{K\hbar^2}(\frac{8\pi\gamma
\ell_{pl}^2}{6})^{\frac{3}{2}}}.
\label{Hami -1}
\end{align}

On the other hand, we also use \eqref{k=-1} to obtain the path integral formulation of $k=-1$ model under deparameterized framework as
\begin{align}
&A_{dep}(v_f,\phi_f;v_i,\phi_i)^{k=-1}\nonumber\\
=&c'_{-1}\int\mathcal{D}v\int\mathcal{D}b~\exp\frac{i}{\hbar}\int d\phi(\sqrt{{3\pi G}v^2\left(\sin^2b-\gamma^2V_0^{\frac{2}{3}}\bar{\mu}^2\right)}\hbar-\frac{\hbar\dot{v}b}{2}).
\end{align}
The effective Hamiltonian in the deparameterized framework for $k=-1$ FRW model can be derived as
\begin{align}
\mathcal{H}_{\emph{eff}}^{dep}&= -\sqrt{{3\pi G}v^2\left(\sin^2b-\gamma^2V_0^{\frac{2}{3}}\bar{\mu}^2\right)}\hbar.
\end{align}
Though Poisson bracket we obtain the following effective equations
\begin{subequations}
\begin{align}{}
\dot{v}&=\{v,\mathcal{H}_{\textit{eff}}^{dep}\}=\frac{3\pi G v^2\sin2b}{\sqrt{{3\pi G}v^2\left(\sin^2b-\gamma^2V_0^{\frac{2}{3}}\bar{\mu}^2\right)}},\\
\dot{b}&=\{b,\mathcal{H}_{\textit{eff}}^{dep}\}=-\frac{6\pi Gv\sin^2b-6\times3^{1/3}\pi G\gamma^2V_0^{\frac{2}{3}}v^{\frac{1}{3}}}{\sqrt{{3\pi G}v^2\left(\sin^2b-\gamma^2V_0^{\frac{2}{3}}\bar{\mu}^2\right)}}.
\end{align}
\label{eqs -1}
\end{subequations}
It is easy to check that equations \eqref{eqs -1} are equivalent to the effective equations derived by the Hamiltonian \eqref{tls H} or \eqref{Hami -1}, and the classical big bang singularity is also replaced by a quantum bounce in $k=-1$ LQC.

In conclusion, a multiple group-averaging approach is proposed to derive the path integral formulation for all FRW models of $k=0,+1,-1$. In all these models, one can express the transition amplitude in the deparameterized framework in terms of group averaging. The relation between the path integral formulation in timeless framework and that in deparameterized framework is then clarified. The effective Hamiltonian in deparameterized framework and the effective Hamiltonian constraint in timeless framework can respectively be derived by the path integrals. They lead to same equations of motion and hence coincide with each other. The effective Hamiltonian constraints given by the canonical theories of the three models are also confirmed by the path integral formulation. Hence the canonical and path integral formulations of LQC coincide.

\section*{ACKNOWLEDGMENTS}
We would like to thank Dah-wei Chiou and Guo Deng for helpful suggestion and discussion. This work is supported by NSFC (No.10975017) and the Fundamental Research Funds for the Central Universities.

\end{document}